\documentclass[12pt,epsfig,citesort,svgnames,dvipsnames]{article}
\usepackage{multicol}
\usepackage[all]{xy}
\usepackage{graphicx}
\usepackage{todonotes}
\usepackage{amsmath}
\usepackage{amssymb}
\usepackage{amsfonts}
\usepackage{amsthm}

\theoremstyle{definition}

\theoremstyle{remark}

\usepackage{amscd}
\usepackage{latexsym}
\usepackage[official]{eurosym}
\usepackage[english]{babel}
\usepackage[T1]{fontenc}
\usepackage[utf8]{inputenc}
\usepackage{authblk}
\usepackage{natbib}
\usepackage{xcolor}
\usepackage{hyperref}
\usepackage[nottoc]{tocbibind}
\hypersetup{
colorlinks=true,
citecolor=WildStrawberry,
anchorcolor=Black,
filecolor=cyan,
menucolor=magenta,
runcolor=cyan,
linkcolor=RoyalBlue,
linktoc=page, 
urlcolor=TealBlue}

\title{\bf Pure Shape Dynamics: General framework}

\author[a]{Tim Koslowski
\thanks{\href{mailto:t.a.koslowski@gmail.com}{t.a.koslowski@gmail.com}}}
\author[b,c]{Pedro Naranjo
\thanks{\href{mailto:pnpfisica@gmail.com}{pnpfisica@gmail.com}}}
\author[b]{Antonio Vassallo
\thanks{\href{mailto:antonio.vassallo1977@gmail.com}{antonio.vassallo1977@gmail.com}}}
\affil[a]{\normalsize University of W\"urzburg, Theoretical Physics II, Campus Hubland Nord, Emil-Hilb-Weg 22, 97074 Würzburg, Germany}
\affil[b]{\normalsize Warsaw University of Technology, Faculty of Administration and Social Sciences, Plac Politechniki 1, 00-661 Warsaw}
\affil[c]{\normalsize University of Warsaw, Faculty of Philosophy, Krakowskie Przedmie\'scie 3, 00-047 Warsaw}

\date{}

\begin{document}

\maketitle

\begin{abstract}
    We put forward a general framework for describing relational physical theories, which we call \emph{Pure Shape Dynamics} (PSD). Elaborating on the original insights brought about by the Shape Dynamics programme, PSD's novel take on relationalism is its insistence on describing any dynamical system by means of the intrinsic geometry of its associated curve in the suitable relational configuration space of the theory, namely shape space, whereby the corresponding equation of state of the curve expresses the ratio of change of one of its geometric degrees of freedom with respect to another one. The mathematical structure underlying the equation of state is a local section over a natural generalisation of the unit tangent bundle, which we call \emph{shape phase space}.\\

\end{abstract}
\tableofcontents

\section{Introduction}{\label{sec:intro}}

One of the main motivations for the development of the Shape Dynamics (SD) programme was the hope that a completely relational description of gravity would be the basis for a new approach to quantum gravity (see \citealp{390}, for a nice introduction to the subject, with an emphasis on conceptual matters, \citealp{514}, for a pedagogical, yet comprehensive, account, and \citealp{528}, for the Hamiltonian version, which renders the mathematical relation between General Relativity and Shape Dynamics transparent). It is the purpose of this series of papers to complete this development. 

Complete relationalism is not implemented in the present formulation of SD (\citealp{529}), which is formulated as a standard canonical system, whose phase space is the cotangent bundle over relational configuration space (so-called \emph{shape space}) with a time-dependent Hamiltonian. The formulation of \citet{529} is an improvement over the original formulation of \citet{83} in terms of best matching in relative configuration space (i.e., with scale) in that it eliminates scale from the description and is thus formulated in shape space (i.e., without scale). Let us consider this system in more detail to explain how it violates the relational first principles of shape dynamics: Let us denote the (by construction dimensionless and scale invariant) shape degrees of freedom by $q^a$ and their canonically conjugate momenta by $p^a$ and the time parameter by $t$. The variables $(p_a,t)$ carry by construction the dimension of an action. The dynamics is generated by a time-dependent dimensionless Hamiltonian $H(q,p;t)$ through the canonical equations of motion
\begin{equation}
 \frac{d\,f(q,p)}{d\,t}\,=\,\left\{f(q,p),H(q,p;t)\right\}.
\end{equation}
This system of equations contains a {\it global} dynamical similarity\footnote{The term `dynamical similarity' refers to the fact that a simultaneous re-scaling of the units of scale, duration and energy (and in some cases coupling constants) with the correct powers maps solutions of the equations of motion onto solutions of the transformed equations of motion. For an expanded account of dynamical similarity, see \citet{733,734}.} under the joint transformation
\begin{equation}
 t \to c\,t\,\textrm{ and } p_a \to c\,p_a,
\end{equation}
where $c$ is a fixed global constant. This dynamical similarity can be used to make the description of the system independent of the units used to describe the $t$ and the $p_a$. In particular at any point (except $t=0$) one can define the units in such a way that initially $t_0:=|t_{initial}|=1$. Moreover, due to dimensional homogeneity, the Hamiltonian only depends on the dimensionless ratios $u_a:=\frac{p_a}{t}$. However, despite the elimination of the unit of $t$ in the initial value problem and the independence of the Hamiltonian of this choice of units, one needs to include the evolution of the parameter $t/t_0$ in the set of the equation of motion to describe the evolution of these initial values. This introduces a preferred parametrization of the dynamical system in terms of the dimensionless parameter $t/t_0$, so although the unit of time is arbitrary, the system fails to be manifestly relational because it possesses a preferred time parametrization.

Pure Shape Dynamics (PSD) aims at a completely relational description of the dynamics of the universe. The idea behind this approach is to disentangle the objective predictions of the dynamical system that is used to describe the universe from non-dynamical reference structures. 
%that are not dynamical degrees of freedom of this dynamical system. 
Let us consider relational configuration space and the curve traced out by a physical system in said space. Then, PSD seeks to describe the dynamics in terms of an equation that expresses how the intrinsic\footnote{In this paper we will call any geometric property of a curve ``intrinsic'' if it can be properly defined for a curve in a Riemannian manifold and is {\it independent} of the parametrization of the curve.} geometric properties of the curve change relative to one another. We call this an {\it equation of state} of the geometry of the curve in shape space, as opposed to ``equation of motion'', in order to emphasise the unparametrized nature of the curve.
%since it determines the ratios of increments of the geometric properties of the curve. 
%this relational description is achieved by describing the dynamics of (a toy model of) the universe in terms of an equation of state of the intrinsic geometry of the curve traced out by the system on relational configuration space (i.e. shape space). 
This enables us to: i) get rid of the reference structures that appear in the definition of units of (local) increments of time, for no parametrization enters the description of the curve, and ii) banish the reference structures that enter into the definition of standard configuration space.

%\section{Relational Considerations}\label{sec:relationalism}

%\todo[inline]{Tim, I'd say there's no need to go through so much detail about ``relationalism'', because readers are likely familiar with the core idea. I'd suggest we keep the \S\ref{sec:intro} for the motivation, along with a schematic description, in \S\ref{sec:relationalism}, of quotient spaces, $\mathcal Q_{ss}\equiv \mathcal Q/\mathcal G$, illustrating the case of the similarity group $\mathcal G=\mathsf{Sim}(3)$ in the $N$-body system, which will let us introduce ``shape space'' rather intuitively. (Antonio and I did this in the chapter 8-)) And then \S \ref{sec:ShapeSpace} will give the bloody details! We may also merge \S\ref{sec:intro} and \S\ref{sec:relationalism}.}

Relationalism is in essence the clean separation of physical law form the operational definition of a physical object. This is due to the fact that physical laws need to be defined for physical objects which in turn are defined by their defining physical properties, i.e., by parts of the laws. 

It is instructive to illustrate the two points above in the simple case of $N$ free Newtonian point particles (see \citealp{729}, and \citealp{728}, for early attempts at providing a relational account of inertial systems; \citealp[see][chap.~3.1]{514}). Newton's first law states that the motion of free point particles in an \emph{inertial} frame is along \emph{straight} lines with \emph{constant} speed. For the sake of argument, let us consider we have a given notion of ``straight line'' at our disposal given by Euclidean geometry, such that we can apply this law in practice if operational definitions of the notions of ``inertial system'' and ``constant speed'' are available. First an inertial system is operationally defined as that in which free point particles move along straight lines with constant speed. At first sight, this definition seems circular and thus empty, but it is not so in the case of $N\ge\,3$ particles: in this case, one can generically use two of the free particles to define the coordinate origin of a frame of reference which fixes the inertial system. Newton's first law then predicts that the remaining particles will move along straight lines with constant speed. Second, we need to define what ``constant'' speed means. For this we need\footnote{\label{foot:FreeRescaling}Strictly speaking, free particles require only a notion of ``physical speed'' due to the symmetry of Newton's equations under the simultaneous rescaling $E\to\lambda^2\,E$, $p_a^I\to\lambda p_a^I$ and $t\mapsto t/\lambda$. This symmetry allows us to absorb the notion of duration, i.e., the units in which one measures $t$ as a rescaling of the total energy $E$, which drops out of the equations that describe the curve in shape space. In this paper, however, we shall consider general particle models in which the ratio between the total energy and the potential energy enters the equations that describe the curve in shape space.} a notion of ``physical distance'' and a notion of ``physical duration'' defined, respectively, by material rods and clocks. This amounts to using one position and one momentum degree of freedom of the remaining particles to define a physical unit of distance and a physical unit of duration. 

Thus, to extract the physical content of Newton's first law, we have to separate the choices of coordinate origin, frame, and total scale---which are defined operationally---from the ``objective'' degrees of freedom, which we will call ``shapes'' for reasons that will become clear shortly, and we also must divide by the choice of unit of time. The space of all physically allowed shapes is thus the configuration space divided by translations, rotations and global scale transformations and is, by construction, independent of the origin, frame, and unit of distance. As emphasised above, independence of the unit of time is achieved by restricting ourselves to unparametrized curves in shape space. 

To see why we call the objective degrees of freedom of a physical system ``shapes'', let us consider 3 particles in Newtonian space, which may be thought of as placed in the vertices of a triangle. Let $x_a$ be the particle positions and $r_{ab}\equiv|x_a-x_b|$ the inter-particle separations, to comply with translational invariance. Further, scale invariance demands that we should take one of the $r_{ab}$ as a unit of length and compare the remaining two against it, yielding the two ratios which are objective. The upshot of this procedure is the fact that only the shape of the triangle matters, for only two independent angles are needed. 

Mathematically, the procedure of systematically getting rid of redundant structure associated with some symmetry is commonly known as \emph{quotienting out} (see \citealp{446}, for a thorough analysis of configuration spaces and their quotient). Schematically, if $\mathcal{Q}$ is the relevant configuration space of a given system and $\mathcal{G}$ is the symmetry group, the quotienting out yielding the space carrying truly physical information is $\mathcal{Q}_{ss}:=\mathcal{Q}/\mathcal{G}$, whose dimension is simply $\mathrm{dim} (\mathcal{Q}_{ss})=\mathrm{dim}(\mathcal{Q})-\mathrm{dim}(\mathcal{G})$. In the case of $N$ classical particles, $\mathcal{Q}$ is standard configuration space, $\mathcal{G}=\mathsf{Sim(3)}$, the joint group of Euclidean translations, rotations and dilatations (called \emph{similarity group}), and $\mathcal{Q}_{ss}=\mathcal{Q}/\mathsf{Sim(3)}$ is the shape space of the system. It is easy to see why such a reduced configuration space is called like this: In this simple scenario, the objective spatial information of the $N$-particle system is encoded in the shape of the $N$-gon defined by the $N$ particles. Hence, in this case, each point on shape space represents a $N$-gon's shape (i.e., an equivalence class of $N$-gons under the joint action of translations, rotations and scaling transformations).

Finally, we should like to emphasise a key aspect of the framework: PSD's take on relationalism accords fundamental status to spatial degrees of freedom and their intrinsic change, with temporal notions, like duration, being effectively derived ones. Thus, the equation of state of the unparametrized curve in shape space explicitly shows the intrinsic change in configurations, $dq^a$, whereas momenta are given implicitly through the unit tangent vector, $u^a$, to the curve at $q^a$. As such, PSD generates the dynamical curve traced out by a given physical system in shape space by means of an initial-value formulation, with initial data \{$q^a, \alpha ^a$\}, where $\alpha ^a$ are the relevant geometric properties of the curve ($u^a$ being one among them). Note that this initial-value formulation is \emph{not} Hamiltonian, for the RHS of the equation of state does not exhibit any Hamiltonian function. However, we do nonetheless make use of Hamilton's equations when computing the equation of state. We shall elaborate on this in \S \ref{subsec:EquationState}.

\section{Shape Phase Space\label{sec:PhaseSpace}}

%\todo[inline]{Forgot to mention I'd just polished a bit the text of this section, in order to make the transition to \S\ref{sec:EqState} smooth. Otherwise, all details are unchanged.}

This section will be concerned with the $E=0\,\,\,N$-body system. See comments at the end of \S\ref{subsec:directional} regarding the $E\ne 0$ case. Suitable generalisations to the case of dynamical geometry and the quantum theory are work in progress.

The key mathematical structure underlying the equation of state of the geometry of the curve in shape space---which guarantees the unparametrized character of said equation---is the canonical action of a local section over a special kind of fibre bundle structure. We shall illustrate the construction of this structure in the following subsections, which will lead to the notion of \emph{shape phase space}. 

\subsection{Unit Tangent-Acceleration Bundle}\label{subsec:UTA}

%\todo[inline]{Tim, how general is $UTA(\mathbb X)$? I mean, it seems that this construction is tailored for the $E=0\, N$-body system, where the $k^a$ satisfy the constraints \eqref{equ:AccelerationConstraint}. I mean, already the $E\ne 0\,N$-body system goes beyond $\{q, \phi , K\}$, with an additional degree of freedom being required ($\varepsilon$ in our draft). And what about ADM gravity? Or Bianchi IX, for that matter (with its $\sigma$ parameter)? Is $UTA(\mathfrak S)$ robust enough to accommodate these generalisations? I mean, knowing the structural similarities between the $N$-body system and GR, I guess it's safe to say there must exist suitable generalisations of $UTA(\mathfrak S)$ to GR and, hence, to realistic cosmological models. But I feel this needs to be spelled out, if only schematically, in this paper. Once we resume the PSD version of GR we'll tackle the issue, and perhaps build its associated unit tangent bundle structure from scratch 8-)}

Given a point $z$ with local bundle coordinates $(x,\alpha)$ in the unit tangent bundle over a Riemannian manifold $(\mathbb X,g)$, one can always construct an arc-length parametrized $g$-geodesic $h_z(s)$ that satisfies $$\varphi_i(h_z(s=0))=x\quad\textrm{ and }\quad 
\left.\left(\frac{\partial \varphi_i(\gamma(s))}{\partial s}\right)\right| _{s=0}=\alpha\,,$$

where $\varphi _i$ is the function $\varphi:U_i\rightarrow\mathbb R$ in the coordinate chart $(U_i,\varphi _i)$ and $\gamma(s)$ is an arbitrary arc-length parametrized curve. Thus, we have a $g$-geodesic $h_z(s)$ that is tangent to $\gamma(s)$ at $x$ with direction determined by $\alpha$. Using these geodesics we find the covariant derivative of $\gamma (s)$ as
\begin{equation}
  \frac{D^2\gamma(s)}{Ds^2}=\frac{d}{ds}\left(\gamma^\prime(s)-h^\prime_{z(\gamma,s)}(s)\right),
\end{equation}
where $z(\gamma,s)$ is the point in the unit tangent bundle that represents the point $\gamma(s)$ and the direction defined by $\varphi _i ^{\prime}(\gamma (s=0))$. $D^2\gamma/Ds^2$ is the directional covariant derivative of the unit tangent vector $u_\gamma$ of the curve in the direction of the unit tangent vector, which is usually denoted by $\nabla^{(g)}_u u$. This allows us to construct the unit tangent-acceleration bundle $UTA(\mathbb X)$ over the base manifold $\mathbb X$ starting from an atlas $\{(U_i,\varphi _i)\}$ for $\mathbb X$, from which we get the union
\begin{equation}
  UA\mathbb U:=\cup_{i=1}^k(\varphi _i(U_i)\times\mathbb S^{n-1}\times\mathbb R^n).
\end{equation}
We then define the equivalence relation $\sim$ as follows:  $(x,\alpha,k)\in \varphi _i(U_i)\times \mathbb S^{n-1}\times \mathbb R^n$ and $(y,\beta,l)\in\varphi _j(U_j)\times \mathbb S^{n-1}\times \mathbb R^n$ are equivalent if there exists an arc-length parametrized curve $\gamma(s)$ with $\varphi _i(\gamma(s=0))=x$ and $\varphi _j(\gamma(s=0))=y$ and 
\begin{equation}
  \left.\left(\frac{\partial\varphi _i(\gamma(s))}{\partial s}\right)\right| _{s=0}=\alpha
\quad\textrm{ and }\quad 
\left.\left(\frac{\partial\varphi _j(\gamma(s))}{\partial s}\right)\right| _{s=0}=\beta\,,
\end{equation}
as well as
\begin{equation}
  \varphi _i^\prime\left(\frac{D^2\gamma(s=0)}{Ds^2}\right)=k\quad\textrm{ and }\quad \varphi_j^\prime\left(\frac{D^2\gamma(s=0)}{Ds^2}\right)=l\,,
\end{equation}
where $\varphi _i^\prime$ and $\varphi^\prime_j$ denote the derivative map of the coordinate functions, as needed for the tangent vectors $k$ and $l$. The unit tangent-acceleration bundle is then defined as $UTA(\mathbb X):=UA\mathbb U/\sim$. Notice that the equivalence relation $\sim$ identifies open sets with open sets, so the quotient inherits the local product space topology. Physically, for $\mathbb X=\mathfrak S$, $UTA(\mathfrak S)$ corresponds to the space of local intrinsic geometric data of curves in shape space $\mathfrak S$, which will be key to expressing the equation of state of the curve in a manifestly unparametrized fashion. Before doing that, however, we need to introduce a second crucial element.

\subsection{Directional Information}\label{subsec:directional}

Let us proceed next to the construction of the canonical action, which underlies the equation of state of the geometry of the curve in shape space.

It is well-known that there exists a canonical action on a unit tangent bundle. In a similar fashion, it is possible to identify the canonical action on the unit tangent-acceleration bundle $UTA(\mathbb X)$. For this, we consider the unit tangent bundle over $UTA(\mathbb X)$: given local coordinates $x^a$ on the base manifold $\mathbb X$ and a local coordinate basis $\partial_{x^a}$ for the tangent vectors, a point $z$ in $UTA(\mathbb X)$ with base point $x\in \mathbb X$, direction $\alpha$ and acceleration $k^a$, the canonical action $A_{z}$ satisfies\footnote{Notice that the canonical action $A_z$ denotes neither an action functional nor a group action. Rather, the term action refers to the generalisation of the fact that a tangent vector at a point acts canonically as a directional derivative on smooth functions with support in a neighbourhood of that point. Similarly, a unit tangent vector defines a ratio of the directional derivatives of two functions independently of the metric that has been used to construct representative unit norm vector.}
\begin{equation}\label{equ:CanonicalActionOnP}
   A_z(x^a,x^b)=\frac{u^a(\alpha)}{u^b(\alpha)},\,A_z(x^a,u^b)=\frac{u^a}{k^b},\, A_z(u^a,x^b)=\frac{k^a}{u^b},\, A_z(u^a,u^b)=\frac{k^a}{k^b}\,.
\end{equation}
This canonical action encodes the directional derivative and the covariant arc-length derivative along the curve and are given by the geometry of the curve itself. We note that the action $A_z$ satisfies by definition
\begin{equation}\label{equ:ActionTransitivity}
    A_z(f,g)A_z(g,h)=A_z(f,h)
\end{equation}
for any triple of functions $f,g,h$ on $UTA(\mathbb X)$. Moreover, since $A_z$ is a derivative action on $UTA(\mathbb X)$, it is linear:
\begin{equation}
    A_z(\lambda_1\,f_1+\lambda_2\,f_2,g)=\lambda_1\,A_z(f_1,g)+\lambda_2\,A_z(f_2,g)
\end{equation}
for any triple $f_1,f_2,g$ of functions on $UTA(\mathbb X)$ and any pair $\lambda_1,\lambda_2$ of real numbers. Furthermore, the action satisfies the Leibniz rule
\begin{equation}
    A_z(f_1\,f_2,g)=f_1\,A_z(f_2,g)+f_2\,A_z(f_1,g)
\end{equation}
for any triple $f_1,f_2,g$ on $UTA(\mathbb X)$. 

Remarkably, the ratios of changes of local coordinate functions completely encode the direction. Hence, we can uniquely specify a local section on the unit tangent bundle over $UTA(\mathbb X)$ by specifying the action $A_z$ for the local coordinate functions. We will use this in the following to define a local section.

Equipped with this machinery, we are ready to provide a formal characterization of the equation of state. As already emphasised several times above, the dynamical content of PSD is expressed as the equation of state of the geometric properties of a curve in shape space. This equation of state can be written as the directional action of a local section in the unit tangent bundle over $UTA(\mathfrak S)$, \eqref{equ:CanonicalActionOnP}. For computational ease, it is advisable to work out the equation of state using a suitable metric on shape space as well as an intrinsic parametrization, like arc-length. This shall be explicitly given for the $E=0$ $N$-body system in \S\ref{sec:EqState}, while, for completeness, \S\ref{sec:appEqState} will provide the directional action counterpart, which explicitly shows the unparametrized nature of the equation of state.

We should like to point out that $UTA(\mathfrak S)$ contains superfluous initial data, for the acceleration vectors satisfy constraints, \eqref{equ:AccelerationConstraint} (see \S\ref{sec:appEqState} for details). This leads us to the notion of \emph{shape phase space} $\mathfrak P$, which is defined as the smallest space $\mathfrak P\subset UTA(\mathfrak S)$ on which the equation of state of the geometry of the curve can be written as the directional action of a local section in the unit tangent bundle over $\mathfrak P$. In \S\ref{sec:PhaseSpaceApp}, we shall give the shape phase spaces associated with free dynamics and the $E=0\,\,\,N$-body system.

The construction presented here and in \S\ref{sec:appEqState} and \S\ref{sec:PhaseSpaceApp} can be readily extended to the Newtonian $E\ne 0$ $N$-body system. In this case, one has to include the jerk (third covariant derivative) in the fibres over each point in shape space and thus construct the unit-tangent-acceleration-jerk-bundle along the lines in which $UTA(\mathfrak S)$ is constructed. It then turns out that the acceleration and the jerk data are constrained in an equation analogous to \eqref{equ:AccelerationConstraint}. When solving the constraints, one finds that only one degree of freedom of the jerks is unconstrained, so shape phase space contains one more local coordinate in the $E\ne 0$ case. 

\section{Equation of State of the curve in Shape Space}\label{sec:EqState}

After the general discussion of the mathematical structure needed to manifestly render the equation of state of the curve independent of any parametrization (intrinsic or otherwise), in this section we shall implement the basic premise of PSD in the Newtonian $N$-body system\footnote{For references on the $N$-body problem as formulated in the `scaffolding' of an inertial frame, external clock and reference scale, see \cite{730}, \cite{731}, and \cite{732}.}: formulate a dynamical system for the unparametrized curve in the associated shape space, with the condition that said curve coincide locally with the projection of solutions of the standard Newtonian $N$-body system onto unparametrized curves in shape space.

\subsection{Setup}\label{subsec:setup}

%\todo[inline]{See box at beginning of \S\ref{sec:ShapeSpace}.}

We denote the coordinates of $N \ge 3$ point particles in $\mathbb R^3$ by $r_i^\alpha:\alpha=1,2,3;i=1,...,N$ and their canonically conjugate momenta by $k^i_\alpha:\alpha =1,2,3;i=1,...,N$. The dynamics of the point particles is generated by the standard Hamiltonian $H_0=\sum_{i=1}^N\frac{\sum_{\alpha=1}^3 (k^i_\alpha)^2}{2\,m_i}+V_0(r)$ with potential $V_0(r_1^1,...,r_N^3)$. Using the mean mass $M:=\frac{1}{N}\sum_{i=1}^N m_i$ and the dimensionless mass ratios $\mu_i:=\frac{m_i}{M}$, we can write the kinetic energy as $T=\frac{1}{2M} g_{ij}^{\alpha\beta}\,k_\alpha^ik_\beta^j$, using the configuration space metric $g_{ij}^{\alpha\beta}=\frac{1}{\mu_i}\delta_{ij}\delta^{\alpha\beta}$ whose entries are by construction dimensionless. We now condense the indices to $I(\alpha,i):=3\,i+\alpha$, so we can write the Hamiltonian in the form
\begin{equation}\label{equ:NewtonHamiltonian}
 H_0=\frac{1}{2\,M}\,g^{IJ}\,k_Ik_J+V_0(r^I).
\end{equation}
Our goal is to describe the equation of state of the curve in shape space that is traced out by the $N$-particle dynamics. 
%Shape space is obtained as the quotient under translations, rotations and dilatations of configuration space $\mathbb R^{3N}\setminus \{\textrm{total collision}\}$. 
From the relational perspective, the most important part is the removal of both an absolute unit of size (obtained through quotienting by dilatations) and an absolute unit of duration (obtained by using an intrinsic parametrization for the curve in shape space). We will therefore explicitly factor by dilatations and express the system in an intrinsic parametrization, while we will treat translations and rotations as pure gauge, implemented through the first class constraints
\begin{eqnarray}
 T_\mu =& t_\mu^I\,k_I&\approx 0\,,\\
 R_\mu =&{r_{\mu I}}^J\,r^I\,k_J&\approx 0\,, \label{equ:RoatationConstraint}
\end{eqnarray} 
where $t_\mu^{I(\alpha,i)}=\delta^\alpha_\mu$ and ${r_{\mu\, I(\alpha,i)}}^{J(\beta,j)}=\varepsilon_{\mu\alpha\beta}\delta_{ij}$. In other words, we will describe a gauged equation of state of a curve in \emph{pre}-shape space (the space obtained by quotienting only by translations and dilatations), rather than in shape space. It turns out to be extremely useful to treat rotations as gauge, because this circumvents a number of technical difficulties with their quotient. 

Let us describe the system in more detail before we describe the curve in pre-shape space. First of all, the dynamics is contained in a fixed energy $E$ surface $H=E$, which we can absorb into the potential by replacing $\tilde V := V_0-E$. We can thus express the energy constraint surface as $\chi=H=T+\tilde V\approx 0$. This constraint is equivalent to 
\begin{equation}
  \chi=\frac{1}{2}\,g^{IJ}\,k_Ik_J+M\,\tilde V(r^I)\approx 0\,,
\end{equation}
so we will from now on use $V(r^I):=M\,\tilde V(r^I)$ to absorb the overall mass scale in the coupling constant of the potential.

Moreover, to ensure that translations and rotations are indeed first class constraints we need to impose that the potential $V_0$ be translation- and rotation- invariant. Now, we will explicitly perform the quotient by translations, which is achieved by going to Jacobi coordinates and eliminating the $N$-th Jacobi coordinate $R^a$ (which represents the centre of mass of the system). Jacobi coordinates are generated by the canonical transformation
\begin{equation}
  F=\pi^I_a\,\rho^a_I(r)+P_a\,R^a(r)\,,
\end{equation}
where $I=1,...,N-1$ and where $\rho^a_I(r):=\frac{\sum_{K=1}^I m_Kr^a_K}{\sum_{K=1}^I m_K}-r_{I+1}^a$ and $R^a(r):=\frac{\sum_{K=1}^Nm_Kr^a_K}{\sum_{K=1}^Nm_K}$. The relation between the momenta $k_a^I$ and the Jacobi momenta $\pi_a^I$ and $P_a$ is
\begin{equation}
  k^I_a=\left.P_a\frac{m_I}{M_N}+\sum_{J=I}^{N-1}\frac{m_I}{M_J}\pi^J_a-\pi^{I-1}_a\right|_{\pi^0_a\equiv 0}\,,
\end{equation}
where $M_I:=\sum_{K=1}^I m_K$. The upshot of using Jacobi coordinates is that it keeps the kinematic metric diagonal, the explicit form of which being
\begin{equation}
  T=\frac{(\vec P)^2}{2M_N}+\sum_{I=1}^{N-1}\frac{ (\vec\pi^I)^2}{2M_{I+1}}\,.
\end{equation}
We now set $P_a\equiv 0$ and fix $R^a\approx 0$ using translation invariance. This implies that the structure of the equation does not change when using Jacobi coordinates --the only thing that does change is that the particle index now has range $I=1,...,N-1$. In particular, the energy conservation constraint still takes the form
\begin{equation}\label{equ:ConstraintJacobi}
  \chi _0=\frac{1}{2}\tilde g^{IJ}_{ab} \pi^a_I\pi^b_J+V(\rho)\approx 0\,,
\end{equation}
with a metric that is diagonal in the index pair $(I,a)$ and whose components are explicitly dimensionless mass ratios after the total mass has been absorbed in a redefinition of the potential, as done before.

We will thus from now on assume that we have introduced Jacobi coordinates and (in a slight change of notation) denote them by $r^I, k_J$ with consolidated indices, as we have done for the original particle coordinates --with the restriction that the consolidated indices now have range $I=1,...,3N-3$. In particular, we shall write the constraint (\ref{equ:ConstraintJacobi}) as
\begin{equation}
  \chi _0=\frac 1 2 g^{IJ} k_Ik_J+V(r)\,,
\label{EnergyconstraintFinal}
\end{equation} 
where $g^{IJ}$ is diagonal and its components are dimensionless mass ratios.

\subsection{Separation of shape and scale degrees of freedom}\label{subsec:separationshape}

The simplest way to derive the decoupling of the shape evolution from the scale and duration degrees of freedom is to perform a canonical transformation that separates the overall size of the system. To do this we consider the generating function
\begin{equation}\label{equ:CanonicalTrf}
  F=\frac 1 2 D\,\ln\left(g_{IJ}r^Ir^J/R_0^2\right)+p_a\,q^a(r^I),
\end{equation}
where the functions $q^a(r^I)$ are assumed to be scale invariant, i.e., $r^I\frac{\partial q^a(r^J)}{\partial r^I}=0$, which ensures that the pre-shape index $a$ has range $a=1,...,3N-4$, 
$(p_a, D)$ denote the transformed momenta defined by \eqref{equ:CanonicalTrf} (see also \eqref{canonicalt}) and $R_0$ is a fixed reference scale.

%\todo[inline]{The dilatational momentum in \eqref{equ:CanonicalTrf} hasn't been defined so far.}

%An explicit choice of pre-shape coordinates are the angles $\varphi^a$ on the $3N-4$ dimensional sphere:
%\begin{equation}\label{equ:PreShapeCoord}
 % \begin{array}{rcl}
  %r^1&=&R\,\cos\varphi^1\\
  %r^I&=&R\,\sin\varphi^1\,...\,\sin\varphi^{I-1}\,\cos\varphi^I\\
  %r^{3N-3}&=&R\,\sin\varphi^1\,...\,\sin\varphi^{3N-4}\,,
  %\end{array}
%\end{equation}
%where $R:=R_0\,e^X:=\sqrt{g_{IJ}r^Ir^J}$. We will mostly work with general pre-shape coordinates $q^a$, but refer to the $\varphi^a$ as a reference choice whenever we use explicit coordinates.

Given the canonically conjugate pairs $\{X,D\}$, $\{q^a,p_a\}$ and $\{r^I,k_I\}$, the canonical transformation generated by (\ref{equ:CanonicalTrf}) is
\begin{equation}
 \begin{array}{rcl}
   X&=&\frac 1 2\,\ln\left(g_{IJ}r^Ir^J/R_0^2\right)\\
   q^a&=&q^a(r^I)\\
   k_I&=&\frac{D}{R_0^2}e^{-2X}g_{IJ}r^J+p_a\frac{\partial q^a}{\partial r^I}\,,
 \end{array} 
 \label{canonicalt}
\end{equation}
the first of which allows us to define $R:=R_0\,e^X=\sqrt{g_{IJ}r^Ir^J}$. Finally, \eqref{equ:CanonicalTrf} transforms the Hamiltonian into
\begin{equation}\label{equ:HamOnPreShapeSpace}
 H=\frac{1}{2R^2}\left(D^2+ g^{ab}(q)p_ap_b\right)+V(R,q)\,,
\end{equation}
where we used the kinematic metric on pre-shape space defined by the equation
\begin{equation}\label{equ:kinematicMetric}
\frac{1}{R^2}\,g^{ab}(q):=
g^{IJ}\frac{\partial q^a}{\partial r^I}\frac{\partial q^b}{\partial r^J}\,.
\end{equation}
\eqref{equ:HamOnPreShapeSpace} provides at once the physical meaning of the variables involved: $q^a$ are the pre-shape degrees of freedom, $p_a$ their conjugate momenta and $D$ is the so-called \emph{dilatational momentum}\footnote{\label{foot:dilatational}The dilatational momentum is defined as $D\equiv\sum _a^N \mathbf{r}_a^{\mathrm{cm}}\,\cdot \mathbf{p}^a_{\mathrm{cm}}$. Given its monotonicity, the ratio $D/D_0$, with $D_0$ some arbitrary choice, has arguably been used as a physical time variable (\citealp{712,529}).}.

Moreover, the rotation constraint transforms into the form $R_\mu=R_0^2e^{2X}(r^D_\mu(q)D+r^a_\mu(q)p_a)\approx 0$, which is equivalent to $r^D_\mu(q)D+r^a_\mu(q)p_a\approx 0$, since the reference scale $R_0>0$.

We can now discuss the dependence of the potential on the scale degree of freedom $R$. To do this it is useful to rewrite the energy conservation constraint $H=T+V_0-E\approx 0$ as a geodesic constraint on configuration space:
\begin{equation}
  \chi=h^{IJ}(r)k_Ik_J-1\approx 0,
  \label{geodesicConst}
\end{equation}

where we use the complexity metric $h^{IJ}(r):=\frac{1}{2}\frac{g^{IJ}}{E-V_0(r)}$, which reduces to $h^{IJ}(r)=-\frac{1}{2}\frac{g^{IJ}}{V_0(r)}$ for the important case $E=0$. The dependence of $h^{IJ}(R,q)$ on the scale $R$ falls into four distinct classes:
\begin{enumerate}
 \item $h^{IJ}(R,q)$ is independent of $R$, which implies that the energy conservation constraint is equivalent to $\frac{1}{2}(D^2+g^{ab}(q)p_ap_b)+\gamma C(q)\approx 0$, which 
in turn implies that initial data with $D$ has the solution $D\equiv D_{init.},X=X_{initial}+D_{init.}\,t$, so the evolution in the scale and duration degrees of freedom is autonomous and does not interact with the shape degrees of freedom.
 \item $h^{IJ}(R,q)$ depends homogeneously on $R$, i.e., $h^{IJ}(R,q)=R^{k\ne 0}\tilde h^{IJ}(q)$, which means that the energy conservation constraint is equivalent to $\frac{1}{2}(D^2+g^{ab}(q)p_ap_b)+\gamma\,R^{k+2}\,C(q)\approx 0$. In this case the system possesses a dynamical similarity under the simultaneous transformation $(q^a,R,D,p_a)\to(q^a,\lambda^{2/(k+2)} R,\lambda D,\lambda p_a)$. This similarity transformation leaves the curve in pre-shape space invariant and changes only its parametrization. It thus follows that dynamical similarity decouples a scale degree of freedom from the dynamics, by always choosing $\lambda$ such that the respective degree of freedom does not evolve.
 \item The generic case in which $h^{IJ}(R,q)$ is not homogeneous. This case is for example obtained in the $N$-body problem with Newtonian gravitation and non-vanishing energy, where the potential takes the form $V(R,q)=\gamma \,C(q)/R-E$.
 \item Free particles, i.e., a vanishing potential, is an exceptional case that can be seen as a special case of the first and the third ones. Therefore, the decoupling of the dynamics on shape space from all scale degrees takes place as in the first case, despite the non-vanishing energy.
\end{enumerate}  

\subsection{Equation of state}\label{subsec:EquationState}

After the preliminary work, this subsection shall finally give the equation of state of the unparametrized curve in pre-shape space traced out by the $N$-body system. A word is in order. In \S\ref{subsec:directional} we have shown how the equation of state can be cast into a manifestly unparametrized fashion by means of the directional action of $A(q)$. This has the shortcoming of making explicit computations rather involved, obscuring the physical insights. Thus, it proves very useful to employ an intrinsic parametrization, like arc-length, with respect to a given metric on pre-shape space, such as the kinematic metric $g_{ab}(q)$ introduced above. 

Using this kinematic metric one obtains
\begin{equation}
 \left(\frac{ds}{dt}\right)^2=g_{ab}(q)\frac{dq^a}{dt}\frac{dq^b}{dt}=g^{ab}(q)p_ap_b\,.
 \label{arc-length}
\end{equation}

Given \eqref{arc-length}, the algorithm for computing the equation of state is straightforward. We start with considering the intrinsic change of shape $\frac{dq^a}{ds}$ and use the canonical equations of motion generated by the energy conservation constraint function $H=\frac{1}{2R^2} (D^2+g^{ab}(q)p_ap_b)+V(R,q)$. Next, we demand that the right-hand side be described in terms of dimensionless and scale-invariant quantities, \{$q^a\,,\alpha _I^a$\}, with $\alpha _I^a$ being a set of intrinsic geometric properties of the curve. For consistency, in order to ensure that the dynamical system given by the equation of state close, the elements in $\alpha _I^a$ must exhaust the set of all possible dimensionless and scale-invariant quantities that can be formed out of the different parameters entering a given theory. This way, we obtain a system of equations that describes intrinsically how the geometric quantities \{$q^a\,,\alpha _I^a$\} change along the curve in pre-shape space.

The first step leads to the trivial kinematic statement ``the change of pre-shape is in the direction of the change of pre-shape.'' Explicitly:
\begin{equation}\label{equ:KinematicEOM}
 \frac{dq^a}{ds}=g^{ab}(q)\frac{p_b}{\sqrt{g^{cd}(q)p_cp_d}}\,,
\end{equation}
where we identify the RHS as the unit tangent vector (w.r.t. the kinematic metric) defined by the pre-shape momenta $p_a$. We will denote the direction degrees of freedom defined by the pre-shape momenta by $\phi_A$, with $A=1,...,3N-5$, and note that these are homogeneous of degree zero in the $p_a$.

To find an explicit representation of the direction degrees of freedom, we use the fact that $g^{ab}(q)$ is a real symmetric matrix and can thus be diagonalized using an orthogonal transformation $O^a_b(q)$, i.e., there exists an orthogonal transformation $O^a_b(q)$, such that $O^a_c(q)O^b_d(q)g^{cd}(q)=\lambda^a(q)\delta^{ab}$, where no summation over $a$ is implied. Using $\pi^a(q,p):=\sqrt{\lambda_a(q)}\,O^a_b(q)\,p^b$  we define the explicit coordinates for the direction $\phi_A$ in pre-shape space as
\begin{equation} 
 \begin{array}{rcl}
   \pi^1&=&|p|\sin\phi_1...\sin\phi_{3N-4}\\
   \pi^a&=&|p|\sin\phi_1...\sin\phi_{a-1}\cos\phi_a\\
   \pi^{3N-4}&=&|p|\cos\phi_{1}
   \label{picoordinates}
 \end{array}
\end{equation}
The definition of these (or any other set of) coordinates for the direction in pre-shape space defines a set of phase space functions $\Phi_A(q,p)$, which are by construction homogeneous of degree zero in the pre-shape momenta. Moreover, we can use the definition of the coordinates to provide an explicit function of the unit (w.r.t. $g^{ab}(q)$) co-tangent vector $u_a(q,\phi)$ defined by the direction $\phi_A$ at a point $q^a$ in pre-shape space. This allows us to write the first equation of motion as
\begin{equation}
 \frac{dq^a}{ds}=g^{ab}(q)u_b(q,\phi)\,.
 \label{unittangent}
\end{equation}

Using any local coordinates $\phi _A$ on the fibre of the unit tangent bundle, we can derive an explicit expression $\Phi _A(q,p)$ for these fibre coordinates in terms of the canonical variables $(q^a, p_a)$, like the one defined through \eqref{picoordinates}. Given such an expression, we can derive their change along the curve in pre-shape space using the canonical equations of motion generated by the energy conservation constraint function $H$:
\begin{equation}\label{equ:DirectionEOM1}
  \frac{d \phi_A}{ds}=\frac{\partial\Phi_A}{\partial q^a}\,g^{ab}(q)u_b(q,\phi)-\frac{1}{p^2}\frac{\partial \Phi_A}{\partial u_a}\left(R^2\frac{\partial V(R,q)}{\partial q ^a}+\frac 1 2 {g^{bc}}_{,a}(q)\,p_bp_c\right),
\end{equation}
where it follows form the homogeneity of degree zero of the functions $\Phi_A(q,p)$ in the pre-shape momenta $p_a$ that $\frac{\partial \Phi_A}{\partial q^a} (q,\phi)$ as well as $\frac{\partial \Phi_A}{\partial u_a} (q,\phi)$ are {\it independent} of the length of the shape momenta $p:=\sqrt{g^{ab}(q)p_ap_b}$, i.e., they depend only on the direction $\phi_A$. 

Next, we express the entire RHS of equation (\ref{equ:DirectionEOM1}) in terms of dimensionless and scale-invariant quantities. This is the point where the dependence on the scale variable $R$ implies distinct behaviour. The first case is that of a geodesic theory w.r.t. a given metric $g^{ab}(q)$ on pre-shape space (which is not necessarily the kinematic metric). In this case the potential is $1$ and the equations of motion for the direction simplify to $\frac{d\phi_A}{ds}=\frac{\partial \Phi_A}{\partial q^a}(q,\phi)g^{ab}(q)u_b(q,\phi)-\frac{1}{2}\frac{\partial \Phi_A}{\partial u_a}{g^{bc}}_{,a}(q)u_b(q,\phi)u_c(q,\phi)$, which means that the geodesic dynamical system closes at this stage.

The second case discussed above occurs when the potential is homogeneous in the scale degree of freedom $R$, i.e., $V(R,q)=\gamma\,R^k\,C(q)$. In this case one encounters the dimensionless and scale-invariant degree of freedom
\begin{equation}\label{equ:KappaDefi}
 \kappa:=\frac{p^2}{\gamma R^{k+2}}\,,
\end{equation}
which we can use to express the equations of motion for the direction in dimensionless and scale-invariant terms:
\begin{equation}
  \frac{d\phi_A}{ds}=\frac{\partial\Phi_A}{\partial q^a}\,g^{ab}(q)u_b(q,\phi)-\frac{\partial \Phi_A}{\partial u_a}\left(\frac{1}{\kappa}\frac{\partial C(q)}{\partial q ^a}+\frac 1 2 {g^{bc}}_{,a}(q)\,u_b(q,\phi)u_c(q,\phi)\right)\,.
\label{direction}
\end{equation}

%\todo[inline]{Check the scaling of $R^{k+2}$https://www.overleaf.com/project/6194369b83b0be6df2bce600 and the factors, in particular in the eom. for $\kappa$ (it's $k+2$, right?)}

To close the dynamical system one needs an equation of motion for $\kappa$, which we again obtain from the canonical equations of motion:
\begin{equation}
 \frac{d\kappa}{ds}=-(k+2)\kappa\frac{D}{p}-2g^{ab}(q)\frac{\partial C(q)}{\partial q^a}u_b(q,\phi)\,, 
\end{equation}
where the new dimensionless and scale-invariant degree of freedom $\varepsilon := \frac D p$ appears. The dynamical relevance of $\varepsilon$ is the point where the difference between homogeneous and generic potentials appears: In the case of a homogeneous potential, one has already absorbed all dimensionful constants, so the energy conservation constraint $H=0$ can be written in completely dimensionless form using the dimensionless variables $(q^a,\phi_A,\kappa,\varepsilon)$, which can thus be (locally) solved for $\varepsilon(q,\phi,\kappa)$. This means that in the case of a homogeneous potential the dynamical system closes at this stage with the equation of motion for $\kappa$:

\begin{equation}
  \frac{d\kappa}{ds}=-(k+2)\kappa\varepsilon(q,\phi,\kappa)-2g^{ab}(q)\frac{\partial C(q)}{\partial q^a}u_b(q,\phi)\,.
\end{equation}

This decoupling of $\varepsilon$ is a consequence of the dynamical similarity that we described above. 

A generic potential, however, does not possess dynamical similarity and one can thus no longer solve the energy conservation constraint for $\varepsilon$. Hence, one has to derive the equation of motion for $\varepsilon$ itself analogously to the previous steps (below
we will do this explicitly for the Newtonian $N$-body problem with non-vanishing energy.) Including  $\varepsilon$ we have exhausted all independent dimensionless and scale-invariant ratios that we can form from the dynamical variables. We can use the Hamilton constraint to express any dimensionless ratio that we can form out of the remaining coupling constants and the last scale degree of freedom. This implies that we can always close the dynamical system at this step with an equation $\frac{d\varepsilon}{ds}=E(q,\phi,\kappa,\varepsilon,\sigma)$ using the local\footnote{\label{foot:signAmb}There are in general several branches of the solution for $\varepsilon$ and to complete the dynamical system one has to specify which branch to choose when encountering a bifurcation point. This shall be addressed below, \S\ref{subsec:Janus}.} solution $\sigma(q,\phi,\kappa,\varepsilon)$ to the energy conservation constraint. Notice that this is independent of how many dimensionful coupling constants are originally present in the system.

Furthermore, in the most general case, one also finds $\kappa$, but the choice is no longer canonical, because there are several dimensionful coupling constants $\gamma_i$ that can be used to form several distinct $\kappa_i:=\frac{p^2}{\gamma_i\,R^{k_i+2}}$, where one of the $\kappa_j$ is the ratio $\frac{p^2}{R^2\,E}$, which we will call $\sigma^{-1}$ later on. In this case, one needs only one of them to express the RHS of the equation of motion of the direction in terms of manifestly dimensionless and scale-invariant quantities.

%\todo[inline]{Check equations match actual computations}

Finally, let us analyse the $E\ne 0$ case as an explicit application of the remarks of the last two paragraphs. As emphasised above, we have to find the equation of motion for $\varepsilon=\frac{D}{p}$. Using $\frac{d\varepsilon}{ds}=\frac{1}{p}\left(\frac{d\,D}{d\,s}-\varepsilon\frac{d\,p}{d\,s}\right)$ as well as 
$\frac{1}{p}\frac{d\,p}{d\,s}=-\frac{u^a\,C_{,a}}{\kappa}$ and $\frac{1}{p}\,\frac{d\,D}{d\,s}=\frac{C(q)}{\kappa}+\frac{R^2\,E}{p^2}$ and the solution 
\begin{equation}
  \sigma:=\frac{R^2\,E}{p^2}=\frac 1 2(1+\varepsilon^2)-\frac{C(q)}{\kappa} 
\end{equation}
to the energy conservation constraint $0=\frac 1 2(p^2+D^2)-\gamma\,R\,C(q)-R^2\,E$, we find\footnote{Here we have chosen to write Newton potential as $-\gamma \frac{C(q)}{R}$, so we can identify $C(q)$ with the complexity of a shape (see \S\ref{subsec:complexity}).} 
\begin{equation}
  \frac{d\,\varepsilon}{d\,s}=\varepsilon\frac{u^a\,C_{,a}(q)}{\kappa}+\tfrac{1}{2}(\varepsilon^2+1)\,.
\end{equation}
We thus find the equation of state of the curve in pre-shape space that is traced out by a positive energy system to be\footnote{Following the general discussion of the local section in \S\ref{subsec:directional}, its defining expression, \eqref{equ:CanonicalActionOnP}, is shown in \S\ref{sec:appEqState} for the case of the $N$-body problem, yielding \eqref{equ:SectionInUtATAS}, whereby, as stressed in the paragraph following it, the equation of state of the curve in pre-shape space is rendered manifestly unparametrized. For instance, the local section can be written as $A(q^a,q^b)=\frac{dq^a/ds}{dq^b/ds}=\frac{dq^a}{dq^b}$, with the arc-length parameter $s$ cancelling out. This is why, for the sake of simplicity, the arc-length parameter $s$ has been dropped from the several equations of state throughout the main body. However, we should like to emphasise that these equations should not be read independently, but, as per \eqref{equ:SectionInUtATAS}, understood to define ratios of changes of local geometric properties of the curve in pre-shape
space.}
\begin{equation}
\begin{array}{rcl}
   d\,q^a &=& u^a(q,\phi)\\
   d\,\phi_A &=& \frac{\partial \Phi_A}{\partial q^a}\,u^a(q,\phi)-\frac{\partial \Phi_A}{\partial u^a}\left(\frac{C_{,a}(q)}{\kappa}+\frac 1 2 g^{bc}_{,a}(q)u_b(q,\phi)u_{c}(q,\phi)\right)\\
   d\,\kappa &=& -(k+2)\kappa\,\varepsilon-2\,C_{,a}(q)u^a(q,\phi)\\
   d\,\varepsilon &=& \varepsilon\frac{u^a\,C_{,a}(q)}{\kappa}
    +\frac{C(q)}{\kappa}+\sigma\,.
 \end{array}
\end{equation}
Moreover, using that $\frac{R^2}{p^2}> 0$, we find that $\textrm{sgn}(E)=\textrm{sgn}(\sigma)$:
\begin{equation}
 \textrm{sgn}(E)=\textrm{sgn}\left(\frac 1 2(1+\varepsilon^2)-\frac{C(q)}{\kappa} \right)
\end{equation}
which allows us to distinguish positive from negative energy systems by simply looking at the local geometry of the unparametrized curve in pre-shape space. 

We should like to stress that the most interesting case is that with a homogeneous potential, because it is structurally analogous to ADM gravity (see \citealp{712}) for an enlightening analysis). Let us gather the relevant expressions: the equation of state of the curve in pre-shape space traced out by the projection of the canonical equations of motion of the homogeneous $N$-body system onto pre-shape space becomes:
\begin{equation}\label{equ:DynSyst}
 \begin{array}{rcl}
  dq^a&=&g^{ab}(q)u_b(q,\phi)\\
  d\phi_A&=&\frac{\partial\Phi_A}{\partial q^a}\,g^{ab}(q)u_b(q,\phi)-\frac{\partial \Phi_A}{\partial u_a}\left(\frac{1}{\kappa}\frac{\partial C(q)}{\partial q ^a}+\frac 1 2 {g^{bc}}_{,a}(q)\,u_b(q,\phi)u_c(q,\phi)\right)\\
  d\kappa&=&-(k+2)\kappa\varepsilon(q,\phi,\kappa)-2u^a(q,\phi)C_{,a}(q)\,.
 \end{array}
\end{equation}

\subsection{Geometric Interpretation}\label{subsec:geometric}

The ingredients \{$q^a,\phi_A$\} in the dynamical system (\ref{equ:DynSyst}) have a clear geometric interpretation in terms of an unparametrized curve in pre-shape space: the former is a point on the curve, while the latter is the tangent direction at said point. The parameter $\kappa$, on the other hand, does not \emph{a priori} enjoy such a simple interpretation. To obtain a geometric interpretation of $\kappa$, it will prove useful to look at the curvature of a curve with respect to a given metric on pre-shape space.

The notion of the curvature of a curve requires a metric $g_{ab}(q)$, whose geodesics define locally what ``straight'' lines are. The curvature of a curve $\gamma$ at a point $q\in\gamma$ is then a measure of the local deviation of $\gamma$ from the geodesic $h$ that is tangent to $\gamma$ at $q$. Thus, let us consider a point $q^a$ and a tangent direction $\phi_A$ and define the curvature $K_\gamma(q)$ of the curve $\gamma$ at the point $q$ (as measured by the metric $g_{ab}$) as

\begin{equation}
 K^2_\gamma(q):=g_{ab}(q)\left(\frac{d}{ds}\left(u_\gamma^a(q,\phi)-u^a_{geo}(q,\phi)\right)\right)\left(\frac{d}{ds}\left(u_\gamma^a(q,\phi)-u^a_{geo}(q,\phi)\right)\right)\,,
 \label{curvaturedef}
\end{equation}
where $u^a_\gamma(q,\phi)$ denotes the unit (w.r.t. $g_{ab}$) tangent vector of the curve $\gamma$ (cf. \eqref{equ:KinematicEOM}, \eqref{unittangent}), $u^a_{geo}(q,\phi)$ the unit tangent vector (w.r.t. $g_{ab}$) of the geodesic with identical tangent direction $\phi_A$ at $q^a$ as the curve $\gamma$ and $\frac{d}{ds}$ denotes the arc-length derivative (w.r.t. $g_{ab}$). It is readily seen from \eqref{direction} that \eqref{curvaturedef} becomes\footnote{Notice the sign ambiguity in \eqref{curvaturekappa}, which shall also be resolved in \S\ref{subsec:Janus}.}

\begin{equation}
    K^2=g^{ab}(q)\frac{C_{,a}(q)C_{,b}(q)}{\kappa^2}=\frac{M^2(q)}{\kappa^2}\,,
\label{curvaturekappa}
\end{equation}
where $M(q):=\sqrt{g^{ab}(q)C_{,a}(q)C_{,b}(q)}$. The relation (\ref{curvaturekappa}) allows us to eliminate $\kappa$ from the system of equations (\ref{equ:DynSyst}) and hence obtain a dynamical system for purely geometric data $(q^a,\,\phi_A,\,K)$ of the unparametrized curve in pre-shape space. Nevertheless, we see that $\kappa$ can be interpreted as a curvature radius (rescaled by a factor of $M(q)$), and thus itself being an intrinsic geometric quantity of the curve. \\

We thus arrive at the following result for homogeneous systems: \\

{\it ``The pure curve in pre-shape space that is traced out by the $N$-body system is completely described by an equation of state of its intrinsic geometric degrees of freedom, whereby the changes in the pre-shape $q^a$, the tangent direction $\phi_A$ and the curvature $K$ are described as one traverses the curve.''}

%This result can be straightforwardly generalized to general systems by including the ``jerk'' degree of freedom $E:=\frac{d\,K}{ds}$, which has to be included in the equation f state of the system when the degree of freedom $\varepsilon$ is not determined by the energy conservation constraint. 

\subsection{Complexity}\label{subsec:complexity}

Given the insistence on the intrinsic geometric properties of the curve in shape space, the reader may justifiably wonder whether PSD has the resources to effectively account for dynamical evolution, for it is \emph{unparametrized} curves all the physical system traces out in shape space. The challenge is hence to find an intrinsic feature of the system that serves the purpose of a physically meaningful label of change. To this extent, we shall exploit the idea put forward in \cite{706} that a shape contains structure encoded in stable records. Thus, the evolution is towards configurations (i.e., shapes) that maximize the complexity of the system. The result provides the desired ground for introducing a direction of change, which boils down to the direction of accumulation of the above-mentioned stable records. Clearly, the first thing we must provide is a natural definition of complexity, along with a suitable measure of it, which, to comply with the central tenet of PSD, should be given in terms of the intrinsic geometry of the unparametrized curve in shape space.  

%\todo[inline]{To make sense of a measure of complexity and its growth, the latter being somewhat unsatisfactory in JFT}

Given that the only fully worked-out physical system exhibiting generic formation of stable records is the $N$-body system, we shall use it as an example to motivate our approach to dynamics. It is worth pointing out that promising results come from the vacuum Bianchi IX cosmological model, where a natural candidate exists for a measure of shape complexity in geometrodynamics (see \citealp[][section 3.5]{712}). However, the extension of our arguments to (the shape dynamical version of) full general relativity and quantum mechanics is still a work in progress.
For current purposes, complexity is essentially the amount of clustering of a system, with a cluster being a set of particles that stay close relative to the extension of the total system. Next, we demand that the complexity function grow when (i) the number of clusters do, and (ii) the clusters become ever more pronounced, namely when the ratio between the extension of the clusters to the total extension of the system grows.
%\end{enumerate}

First, recall the centre-of-mass moment of inertia:
\begin{equation}
    I_{\mathrm{cm}}=\sum _{a=1}^{N}m_a\,\mathbf{r}_a^{\mathrm{cm}}\cdot\mathbf{r}_a^{\mathrm{cm}}\equiv\sum _{a<b}\frac{m_a\,m_b}{m_{\mathrm{tot}}}r_{ab}^2:=m_{\mathrm{tot}}\,\ell^2_{\mathrm{rms}}\,,
\end{equation}
where $\mathbf{r}_a^{\mathrm{cm}}$ is the position of particle $a$ relative to the centre of mass, $m_{\mathrm{tot}}=\sum _a m_a$, and $r_{ab}=|\mathbf{r}_a-\mathbf{r}_b|$. Next, the Newton potential reads:
\begin{equation}\label{newt}
    -V_{\mathrm{N}}=\sum _{a<b}\frac{m_a\,m_b}{r_{ab}}:=m_{\mathrm{tot}}^2\,\ell_{\mathrm{mhl}}^{-1}\,.
\end{equation}
Finally, these two equations readily yield
\begin{equation}
    \mathsf{Com\,(q)}=-\frac{1}{m_{\mathrm{tot}}^{5/2}}\sqrt{I_{\mathrm{cm}}}\,V_N=\frac{\ell_{\mathrm{rms}}}{\ell_{\mathrm{mhl}}}\,.
    \label{complexity}
\end{equation}

That \eqref{complexity} is indeed a natural candidate for the complexity function is easily seen, for $\ell_{\mathrm{rms}}$ and $\ell_{\mathrm{mhl}}$ account for the greatest and least inter-particle separations, respectively. Thus, their ratio, \eqref{complexity}, measures the extent to which particles are clustered. It is simple to realize that the opposite of \eqref{complexity} is just the shape space version of \eqref{newt}. Hence, apart from the mass factor, $-\mathsf{Com\,(q)}$ is the shape potential $C(q)$ of the $N$-body system.\footnote{To be precise, all PSD models have an associated shape potential, which is related to the measure of complexity defined in each particular context.}

Now, for $E_{\mathrm{cm}}\ge 0$, $I_{\mathrm{cm}}$ is concave upwards as a function of Newtonian time, and its time derivative (essentially, the dilatational momentum $D$) is monotonic, which further implies that $I_{\mathrm{cm}}$ is U-shaped, with a unique minimum at $D=0$ that divides all solutions in half (see, again, \citealp{706}). This is the \emph{Janus Point}.\footnote{This designation was first put forward in \cite{709}.} The complexity function \eqref{complexity} has a minimum near this point and grows in either direction away from it. This automatically defines a time-\emph{asymmetric} dynamics for internal observers, i.e., ones within one of the two branches at either side of the Janus Point. \S\ref{subsec:Janus} provides this analysis in terms of the intrinsic geometric properties of the curve in shape space.

Moreover, the $N$-body system features generic solutions which break up the original system into subsystems, consisting of individual particles and clusters, that become increasingly isolated in the asymptotic regime \citep{717}. Interestingly enough, such almost isolated subsystems will develop approximately conserved charges, namely the energy $E$, linear momentum $\bf P$ and angular momentum $\bf J$, which enable us to define units $X$ of spatial scale and $T$ of duration as:
\begin{equation}\label{equ:SubsystemUnits}
  X^2=\frac{\bf J^2}{\bf P^2}\,,\,\,\textrm{ and }\,\,T^2=\frac{\bf J^2}{E^2}\,.
\end{equation}

%\todo[inline]{Janus point, time-asymmetric solutions, secular growth of complexity, attractors}

This discussion naturally leads to two crucial remarks. First, the ever better isolated subsystems serve as local and \emph{stable} substructures, whose ever better conserved charges give rise to stable \emph{records}. This dynamically defines a direction of increasing complexity, measured by the complexity function \eqref{complexity}, which, we recall, by construction, tends to grow secularly. We submit that this direction of increasing complexity be identified with the so-called \emph{arrow of time} for internal observers. Thus, we arrive at a description of the experienced asymmetry in the passage of time in terms of purely intrinsic properties of the unparametrized curve in shape space. Second, within the dynamically formed subsystems, there are pairs of particles that may function as physical rods and clocks. These are referred to as \emph{Kepler pairs} because their asymptotic dynamics tend to elliptical Keplerian motion. In \S\ref{subsec:EphemerisEquations}, we will illustrate how suitable equations for standard concepts of scale and duration can be given in terms of the intrinsic geometry of the unparametrized curve in shape space.

\subsection{Ephemeris equations}\label{subsec:EphemerisEquations}

%\todo[inline]{Don't have much to say about this :-P Sure, it's in need of some polishing and the like, but right now this is at the very bottom of my concerns!}

Having the description of the dynamics of a given physical system as an equation of state on pre-shape space, the question immediately arises as to how the standard Newtonian notions of scale and duration may be arrived at from this purely relational description. Although it is clear that it will be impossible to get absolute units of scale and duration, since the curve is void of any such structures, one can nonetheless meaningfully ask how definitions of absolute scales evolve. To explain the question, let us consider a curve $\gamma$ in pre-shape space and two points $q^a$ and $q^b$ on it. Next, we will define the total scale $R$ of the system at point $q^b$ to be the unit of size $R_0$ and the total duration between $q^a$ and $q^b$ to be the unit of time $T$. Then, given a third point $q^c$ on $\gamma$, one may ask: what is the total scale $R$ measured in units of $R_0$? Likewise, what is the duration between $q^b$ and $q^c$ in units of $T$? This second question is analogous to that which historically led to the notion of ``ephemeris time'', hence we shall refer to the resulting equations as ``ephemeris scale-'' and ``ephemeris duration-'' equations, respectively (see \citealp{727}, for one of the first discussions about ephemeris time, and \citealp{726}, for an engaging account of both conceptual and historical topics related to time).

It should be emphasised that the ephemeris equations are model-dependent. For illustration purposes, we shall take the homogeneous model analysed in \S\ref{subsec:EquationState}. The generalisation to other models is straightforward. By virtue of \eqref{equ:CanonicalTrf}, the ephemeris scale equation reads
  \begin{equation}
  \frac{d}{d\,s}\log R=\frac{d}{ds}X=\frac{D}{p}=\pm\sqrt{-\left(1+2\frac{C(q)}{\kappa}\right)}\,,
  \label{ephemerisScale}
\end{equation}

where in the firt step we have used $R=R_0\,e^X$ (cf. \eqref{canonicalt}), in the second step both the fact that $\{X,D\}$ are canonically conjugate variables and the arc-length parametrization condition, \eqref{arc-length}, have been invoked and, finally, the ratio $\frac{D}{p}$ follows at once from the energy conservation constraint $H\approx 0$, \eqref{equ:HamOnPreShapeSpace}, for a homogeneous potential.

Notice that the shape potential $C(q)$ in Newtonian gravity is negative definite, which must be taken into account in \eqref{ephemerisScale}.

Likewise, from the definition of the length of the shape momenta, $p^2=g^{ab}p_a p_b$, we have $\frac{d}{ds} p^2=\tfrac{1}{p}(g^{ab}_{,c}p_a p_b p^c +2g^{ab}p_a \tfrac{d}{dt}p_b)=-\tfrac{2}{\kappa}p^2\,u^a\,C_{,a}$, where the arc-length parametrization condition, \eqref{arc-length}, as well as Hamilton's equations have been used. The ephemeris duration equation follows at once:

\begin{equation}
  \frac{d}{d\,s}\log\left(\frac{ds}{dt}\right)=\frac{d}{d\,s}\log p= -\frac{1}{\kappa}u^a(q,\phi)\,C_{,a}(q)\,.
\end{equation}

\subsection{Janus points and repulsors}\label{subsec:Janus}

From the PSD perspective, the analysis about complexity carried out in \S\ref{subsec:complexity} should be grounded on the intrinsic geometric properties of the associated curve in shape space. Thus, it should be possible to identify the Janus point by directly inspecting said geometric properties. Fortunately, as already hinted at above, this is indeed the case.

As analysed above, the Janus point corresponds to vanishing dilatational momentum, $D=0$, which are special points for the dynamics on shape space, for a sign choice must be made (cf. footnote \ref{foot:signAmb}). Although $D$ itself is not visible on shape space, $\frac{D}{p}$ can be read off equation (\ref{ephemerisScale}). Since $p$ is positive at any regular point of the curve, it suffices to inspect where the RHS of equation (\ref{ephemerisScale}) vanishes to find the location of the Janus point. This gives the condition
\begin{equation}
    \kappa=-2\,C(q)\,.
\end{equation}
Hence the Janus point can be read off the curvature of the curve, without knowing either the scale or the parametrization in Newtonian time. 

We can now address the sign ambiguity in the equation (\ref{curvaturekappa}). This can be resolved by demanding that the perceived direction of the arrow of time, as obtained by the inspection of the secular growth of complexity, be the direction of the forward-integration of the equations of motion. In this case, $\frac{D}{p}$ is positive.

We will now argue that many Janus points are associated with repulsors on shape space, which happens generically when the Newtonian equations of motion imply $\dot D>0$ at the Janus point, which is the generic case if the potential depends on $R$. Moreover, there are special cases where one can bound $\dot D>0$ all over phase space---the simplest such case being a potential of homogeneity degree $-1<k<0$ in $R$ with $E \ge 0$ and shape potential $C(q)<0$. $\dot D>0$ then implies that the scale degree of freedom $R$ takes a minimum at the Janus point. The Liouville volume on Newtonian phase space takes the form $R^{n}dR\wedge d^n\textrm{shape}\wedge d^{n+1}\textrm{momenta}$, so the conservation of the Liouville measure implies that the phase space volume $d^n\textrm{shape}\wedge d^{n+1}\textrm{momenta}$ of the degrees of freedom that influence the curve in shape space decreases rapidly when $R$ grows. The Janus point is thus a repulsor for the dynamics on shape space (\citealp{709}, firstly put forward the discussion of attractors in a framework close to ours; also, see \citealp{735,736}, for analyses of attractors in cosmology).

\section{Discussion}\label{sec:discussion}

This is the first paper of an ongoing research programme whose goal is to implement full relational principles. We have argued that PSD addresses this challenge head on: its main ingredient is the equation of state of the corresponding unparametrized curve in shape space, whereby the dynamics is expressed in terms of the relative change of the intrinsic geometric properties of this curve. As already stressed a number of times above, the unparametrized character of the curve allows us to remove the reference structures that appear in the definition of units of local increments of time, thereby dissolving the defect present in the current formulation of Shape Dynamics, as discussed in \S\ref{sec:intro}. 

Mathematically, the structure underlying this equation of state is the directional action of a local section in the unit tangent bundle over shape phase space, $\mathfrak P$ (cf. \S\ref{subsec:directional}). This local section, \eqref{equ:CanonicalActionOnP}, renders the unparametrized nature of the equation of state explicit. In the case analysed in this paper---i.e., that of the $N$-body system--- the associated shape phase space is detailed in \S\ref{sec:PhaseSpaceApp}.  

The immediate question now is whether this framework can be suitably generalised to more realistic theories. Here we should distinguish the classical versus the quantum cases. We think the former poses no serious obstacles: interestingly, the equation of state associated with the Bianchi IX cosmological model is already worked out (\citealp{718}, lays down the preliminary work and, as already mentioned in \S\ref{subsec:complexity}; \citealp{712}, offers promising results regarding a suitable measure of complexity in this model), and the corresponding shape phase space is a straightforward extension of the construction given in \S\ref{subsec:directional}. For pedagogical reasons, we shall present Bianchi IX along with the PSD version of full GR in a forthcoming follow-up paper. Again, we foresee no serious objections to carrying out these generalisations.

As for the implementation of PSD principles in the quantum realm, both conceptual and technical issues present themselves. First, we should address the question regarding the role that the wave function plays. If we remain committed to the intrinsicness emphasised by PSD, the wave function may very well be thought of as yet another theoretical term reducible, in principle, to the geometric properties of the curve in shape space. However, it is worth remarking that this line of reasoning meets several challenges---e.g., how can the seemingly \emph{branching} structure encoded in the wave function be reduced to a \emph{single}, albeit geometrically complex, curve in shape space? Thus, there exists the possibility that the wave function should be granted special status, being an irreducible part of the framework alongside the intrinsic properties of the curve. This in turn begs the question as to how to account for the dynamical evolution of the wave function itself. The analysis of \S\ref{subsec:complexity} strongly suggests that such a dynamical evolution should not be cast in temporal terms, but should be accounted for by the \emph{complexity} of the system. Hence, this complexity---as opposed to any temporal notion---should appear in the PSD counterpart of the Schr\"odinger equation. Both options about the fundamental status of the wave function are being currently explored and shall be presented elsewhere.  

Furthermore, the construction of shape phase space in the classical case---exemplified for the $N$-body problem in \S\ref{subsec:directional}---might not carry over to the quantum realm as smoothly as one may wish. In this regard, it may be useful to focus on a particular quantum theory that shares with PSD the fundamental role accorded to instantaneous \emph{spatial} configurations. The most promising candidate to this role is the de Broglie-Bohm theory, which has been recently shown to be amenable to the quotienting out procedure discussed in \S\ref{sec:intro} (\citealp{468,530}). Nonetheless, the details of this construction still require much work---as does the extension to quantum field theory--- and will be the subject of our future investigations.

\appendix
\section{Equation of state for $E=0\,\,\,N$-body system as directional action}\label{sec:appEqState}

In this appendix, we shall explicitly construct the mathematical structure underlying the equation of state of the unparametrized curve traced out by the $E=0$ Newtonian $N$-body system in pre-shape space, to complete the analysis of \S\ref{subsec:EquationState}, \S\ref{subsec:geometric}. As already discussed in \S\ref{subsec:directional}, this equation of state can be written as the directional action of a local section in the unit tangent bundle over $UTA(\mathfrak S)$, \eqref{equ:CanonicalActionOnP}. 

In order to appreciate the physics underlying the $UTA(\mathfrak S)$ structure, let us rewrite \eqref{equ:DynSyst} as
\begin{equation}\label{equ:EoScomponents}
    dq^a=u^a(q,\phi)\quad\textrm{ and }\quad d\phi_A=\Phi_A(q,\phi,k)\quad\textrm{ and }\quad dk^a=K^a(q,\phi,k)\,,
\end{equation}
where we have introduced the acceleration vectors $k^a=\frac{\partial u^a}{\partial s}$, which, in turn, enable us to define the curvature as $K^2=g_{ab}k^ak^b$.

We will show how the equations (\ref{equ:EoScomponents})---alternatively \eqref{equ:DynSyst}---of the $E=0$ Newtonian $N$-body system can be written as a section in a unit tangent bundle. Since this system is closely related to the simpler problem of motion of free particles, we will consider the latter as preparation. The equation of state of the curve in pre-shape space traced out by free particles are the geodesic equations 
\begin{equation}
    \frac{d\,q^a}{d\,s}=u^a(q,\phi)\quad\textrm{ and }\quad\frac{d\,u^a(\phi)}{ds}=-\Gamma^a_{bc}(q)\,u^b(q,\phi)u^c(q,\phi)\,.
\end{equation}

Fixing local coordinates on the unit tangent bundle and using $g_{ab}(q)u^a(\phi)u^b(\phi)=1$, we can invert the expression $u^a=U^a(q,\phi)$, where $U^a(q,\phi)$ stands for the unit tangent vector pointing in the direction $\phi_A$ obtained in these coordinates, for angle functions $\Phi_A(q,u)$. This allows us to find a representation of the Christoffel connection on the unit tangent bundle as
$$
\begin{array}{rcl}
  \frac{d\,\phi_A}{d\,s}&=&\left.\left(\frac{\partial \Phi_A(q,u)}{\partial q^a}-\frac{\partial \Phi_A(q,u)}{\partial u^b}\Gamma^a_{ac}(q)u^c(q,\phi)\right)\right|_{u=U(q,\phi)}u^a(q,\phi)\\
  &=:&\Gamma_{A\,a}(q,\phi)\,\,u^a(q,\phi)\,.
\end{array}
$$
We see that at any point $z$ in the unit tangent bundle we can choose Riemann normal coordinates, such that $\Gamma_{A\,a}(z)=0$. Using the connection 1-form $\Gamma_{A\,a}(q,\phi)$ and the representation of the unit tangent vectors $u^a(q,\phi)$, we can define the equation of state of the geometry of the free system in terms of a section on the unit tangent bundle over the unit tangent bundle $UT(\mathfrak S)$ over shape space as:
\begin{equation}
    A(q^a,q^b)=\frac{u^a(q,\phi)}{u^b(q,\phi)},\quad A(\phi_A,q^a)=\frac{\Gamma_{A\,b}(q,\phi)\,u^b(q,\phi)}{u^a(q,\phi)}\,.
\end{equation}
Furthermore, by means of the covariant derivative $D/Ds$ we can conveniently express the arc-length parametrized curve traced out by free particles in pre-shape space using the unit tangent-acceleration bundle $UTA(\mathfrak S)$:
\begin{equation}
    \frac{d\,q^a}{d\,s}=u^a,\quad\frac{D\,u^a}{D\,s}=k^a,\quad \frac{Dk^a}{Ds}=0\,,
\end{equation}
to which we have to add the initial value constraint
\begin{equation}
  k^a\equiv 0
\end{equation}
to describe the curve traced out by free particles. Hence, the acceleration vectors $k^a$ vanish along the curve, namely the dynamics decouples from the acceleration degrees of freedom and the system reduces to
\begin{equation}
   \frac{d\,q^a}{d\,s}=u^a,\quad\frac{D\,u^a}{D\,s}=0\,. 
\end{equation}

Likewise, we will be able to describe the dynamical effect of a translation- and rotation- invariant potential. The first effect is a deviation form geodesics
\begin{equation}
    \frac{D\,u^a}{D\,s}=k^a\,.
\end{equation}
Direct calculation, as preformed in section \ref{subsec:EquationState}, shows that the acceleration degrees of freedom $k^a$ are not independent, but instead satisfy
\begin{equation}
    k_a=\frac{\partial C(q)}{\partial q^a}\,\kappa^{-1}\,,
    \label{equ:AccelerationConstraint}
\end{equation}
which enables us to relate the single independent degree of freedom $\kappa$ to the curvature of the curve $K$ (cf. \eqref{curvaturekappa}):
\begin{equation}
    K^2=g^{ab}(q)\frac{C_{,a}(q)C_{,b}(q)}{\kappa^2}=\frac{M^2(q)}{\kappa^2}\,,
%\label{curvaturekappa}
\end{equation}
where $M(q):=\sqrt{g^{ab}(q)C_{,a}(q)C_{,b}(q)}$. From \eqref{curvaturekappa} and \eqref{equ:AccelerationConstraint} we explicitly see that the curvature $K$ of the curve is the only independent degree of freedom of the acceleration vectors $k^a$. It is clear from the construction that equation (\ref{equ:AccelerationConstraint}) is satisfied as an initial value constraint and is preserved by the dynamics. As discussed in \ref{subsec:geometric}, \eqref{equ:DynSyst} and \eqref{curvaturekappa} allows us to express the equation of change of $K$ along the curve, $\frac{dK}{ds}=\mathfrak K(q,\phi,K)$, whose explicit form we do not give, as is not particularly illuminating.

%\begin{equation}
 %   \begin{array}{rcl}
  %       \frac{d K}{d s}
   %      &=& u^a\left(\frac{K\,M_{,a}}{M}-4K\sqrt{\frac{K}{M}}C_{,a}\right)-2K\sqrt{\left|1-2C\sqrt{\frac{K}{M}}\right|}\,=\,\mathfrak K(q,\phi,K)\,.
    %\end{array}
%\end{equation}
Finally, by \eqref{equ:CanonicalActionOnP}, we arrive at the local section $A$ in the unit tangent bundle over $UTA(\mathfrak S)$ that describes the equation of state of the geometry of the curve in pre-shape space:
\begin{equation}\label{equ:SectionInUtATAS}
  A(q^a,q^b)=\frac{u^a}{u^b},\quad A(\phi_A,q^a)=\frac{\Gamma_{A\,b}u^b+\Delta_{A\,b}k^b}{u^a},\quad A(K,q^a)=\frac{\mathfrak K}{u^a}\,,
\end{equation}
where $\Delta_{A\,b}(q,\phi):=\left.\frac{\partial \Phi_A(q,u)}{\partial u^b}\right|_{u=u(q,\phi)}$. 
%and the acceleration satisfies the constraint (\ref{equ:AccelerationConstraint}), which reduces the acceleration degrees of freedom from $k^a$ to the single independent degree of freedom $K$, as stressed above. 
\eqref{equ:SectionInUtATAS} renders the equation of state of the curve traced out by the $E=0$ $N$-body system in pre-shape space manifestly unparametrized, whereby, as already anticipated in \S\ref{sec:intro}, this equation of state measures the relative change of the relevant geometric properties of the curve.

\section{Examples of Shape Phase Space\label{sec:PhaseSpaceApp}}

In \S\ref{sec:appEqState} we have seen that the acceleration vectors $k^a$ satisfy constraints, which means that not all points in $UTA(\mathfrak S)$ are valid initial data for the equation of state of the curve in pre-shape space, leading to shape phase space, as anticipated in \S\ref{subsec:directional}.

It is illuminating to give the shape phase spaces associated with free dynamics, on one hand, and the $E=0\,\,\,N$-body system, on the other.

The first case implies $k^a \equiv 0$. The subspace of $UTA(\mathbb S)$ at which $k^a=0$ can be canonically identified with the unit tangent bundle $UT(\mathfrak S)$ over shape space $\mathfrak S$ by identifying locally $(q^a,\phi_A)\leftrightarrow(q^a,\phi_A,k^a=0)$. This extends to a global identification since the vanishing of a vector is a statement independent of the chart in which it is expressed. We thus find that the shape phase space for free particles is the unit tangent bundle $UT(\mathfrak S)$ over shape space $\mathfrak S$.

Let us now turn to the PSD description of the relational $N$-body system. The constraint (\ref{equ:AccelerationConstraint}) allows us to coordinatize shape phase space locally by identifying the points $(q^a,\phi_A)$ in the unit tangent bundle and non-negative curvatures $K$ through $(q^a,\phi_A,K)\leftrightarrow (q^a,\phi_A,K\nabla^a C)$ in a chart. By the same argument as in the free particle case, we can identify the subspace $K\equiv 0$ with the unit tangent bundle $UT(\mathfrak S)$ over shape space. We can thus construct shape phase space by putting a $\mathbb R^+_0$ fibre over $UT(\mathfrak S)$. This fibre is not just a manifold, but a manifold with boundary, which can be obtained, for instance, as the quotient of $UTA(\mathbb S)$ by rotations of the acceleration vectors $k^a$ in local Riemann normal coordinates constructed at the base point. This construction defines the shape phase space $\mathfrak P$ of the relational $N$-body system. To emphasise that only the curvature degree of freedom $K=|k|$ of the acceleration vectors forms part of $\mathfrak P$ (recall \eqref{equ:AccelerationConstraint}), we will use the shorthand $UT|A|(\mathfrak S)$ for the shape phase space, which we will call the unit tangent-isotropic acceleration bundle over shape space.

The fact that the fibre possesses a boundary at $K=0$ implies that we have to carefully construct the directional action on the unit tangent bundle over $UT|A|(\mathfrak S)$. It follows from continuity of the equations of motion in $k^a$ that the action has to be repulsive at $K=0$, which means that $A(K,q^a)=\frac{\mathfrak K}{u^a}$ changes sign to $A(K,q^a)=-\frac{\mathfrak K}{u^a}$ at $K=0$. This sign then remains until the next approach to $K=0$. Thus, we have to update the equation of state of the geometry of the curve in pre-shape space by changing the last expression in equation (\ref{equ:SectionInUtATAS}) to
\begin{equation}
  A(K,q^a)=s\,\frac{\mathfrak K}{u^a}\,,
\end{equation}
where the sign $s=\pm 1$ changes at every approach to $K=0$.

\section*{Acknowledgements}\pdfbookmark[1]{Acknowledgements}{acknowledgements}
We are grateful to two anonymous referees for their enlightening comments on a previous version of this paper. Tim Koslowski would like to thank many collaborators for their support of the Shape Dynamics programme. His first and foremost thanks go to Julian Barbour, who has developed many of the conceptual ideas exploited in this research field. Many thanks go also to Henrique Gomes, Flavio Mercati, David Sloan and Sean Gryb for many stimulating discussions about Pure Shape Dynamics. Pedro Naranjo would also like to thank Julian Barbour for his hospitality at College Farm, where P. N.'s interests in relational physics started to develop. Also, discussions with Sean Gryb and Flavio Mercati are much appreciated. P. N. and Antonio Vassallo acknowledge financial support from the Polish National Science Centre, grant nr. 2019/33/B/HS1/01772. 

\bibliography{biblio}

\end{document}